\documentclass[prb,twocolumn,superscriptaddress,showpacs,amsmath,amssymb]{revtex4}
\usepackage{amsfonts}
\usepackage{bbm}
\usepackage{graphicx}% Include figure files
\usepackage{dcolumn}% Align table columns on decimal point
\usepackage{bm}% bold math
\usepackage{subfigure}
\usepackage{multirow}
\usepackage{color}

\begin{document}
\title{Orientation-dependent Josephson effect in spin-singlet superconductor/altermagnet/spin-triplet superconductor junctions}
\author{Qiang Cheng}
\email[]{chengqiang07@mails.ucas.ac.cn}
\affiliation{School of Science, Qingdao University of Technology, Qingdao, Shandong 266520, China}
\affiliation{International Center for Quantum Materials, School of Physics, Peking University, Beijing 100871, China}

\author{Qing-Feng Sun}
\email[]{sunqf@pku.edu.cn}
\affiliation{International Center for Quantum Materials, School of Physics, Peking University, Beijing 100871, China}
\affiliation{Hefei National Laboratory, Hefei 230088, China}
\affiliation{CAS Center for Excellence in Topological Quantum Computation, University of Chinese Academy of Sciences, Beijing 100190, China}

\begin{abstract}
We study the Josephson effect in the spin-singlet superconductor/altermagnet/spin-triplet superconductor junctions
using the Green's function method.
The current-phase difference relationships in the junctions strongly
depend on the orientation of altermagnet and the types of the Cooper pairs.
For the orientation angle equal to odd multiples of $\pi/4$,
the current-phase difference relationships are of the $\sin{2\phi}$ type,
which are irrespective of the pairing wave functions in superconductors.
For the other orientation angles, the emergence of the lowest order current becomes possible and its form, $\sin\phi$ or $\cos\phi$, depends on the pairing wave functions in superconductors.
The $\phi_{0}$ phase and the $0$-$\pi$ transition can be realized in our junctions due to the appearance of the lowest order current.
The selection rules for the lowest order current are presented.
The symmetric relations satisfied by the current-phase difference relationships are analyzed through considering the transformations of the junctions under the mirror reflection, the time-reversal and the spin rotation operations.
Our results not only provide a method to detect the intrinsic spin-triplet superconductivity but also possess application values in the design of the field-free quantum devices.
\end{abstract}
\maketitle

\section{\label{sec1}Introduction}

The third magnetic phase, dubbed altermagnetism\cite{Smejkal1,Mazin1},
has recently attracted great research interest in condensed matter physics\cite{Smejkal2,Mazin2,Turek}. Altermagnet (AM) has the vanishing net magnetism similar to antiferromagnet due to the alternating order of the magnetic moments in the direct space.
However, it also possesses the alternating spin-polarized order in the momentum space, which breaks the time-reversal symmetry just as ferromagnet does. The unconventional anisotropic $d$-wave magnetism can be realized in AM as the magnetic counterpart of the unconventional anisotropic $d$-wave superconductor\cite{Tsuei}.
AM exhibits many novel physical properties and broad applications in various fields such as spintronics, spin caloritronics and superconductivity \cite{Smejkal1,Smejkal3,Bai}.

Because that AM possesses the zero net macroscopic magnetization,
it is of a great benefit in constructing superconducting heterojunctions.
Very recently, the researches on the interplay between altermagnetism and superconductivity are an important aspect, which include the coexistence of altermagnetism and superconductivity and the transport in junctions consisting of AM and superconductor.
$\check{\text{S}}$mejkal $\emph{et al}$.\cite{Smejkal1} and Mazin\cite{Mazin3} have analyzed the possible form of Cooper pairs hosted
in the materials with the altermagnetic order as well as its difference
from that in the ferromagnetic and antiferromagnetic materials.
In the AM/superconductor junction, the effects of altermagnetism on the Andreev reflection are clarified\cite{Sun,Papaj}.
It is found that the charge and spin conductances strongly depend on the interfacial orientation of AM relative to superconductor.
In the superconductor/AM/superconductor junctions, the decay length and the oscillation period of supercurrent show qualitative difference from the ferromagnetic Josephson junctions\cite{Ouassou}. They strongly depend on the crystallographic orientation of AM.
The Andreev levels in the AM Josephson junctions are also studied by Beenakker and Vakhtel\cite{Beenakker}, where the phase difference $\phi$ dependence of the levels is shifted by an offset. The authors demonstrate the non-sinusoidal current-phase difference relationship (CPR) and the $0$-$\pi$ transition in the junctions based on the non-perturbative approach. Recently, the topological superconductivity and Majorana corner states are found in altermagnet with the Rashba spin-orbit coupling or the heterostructure composed of altermagnet and the two-dimensional topological insulator\cite{Zhu,YXLi}.

However, the interlay of altermagnetism and the unconventional spin-triplet superconductivity in the Josephson junctions has not been considered. Josephson junctions are an effective tool to detect the paring symmetry in the unconventional superconductor\cite{Tsuei,Harlingen}. The minimal Josephson model including the spin-triplet superconductor (TS) is the SS/TS junction, which plays an important role in the determination of the order parameter in the candidate materials for TS\cite{Nelson,Zutic,Asano2005,Kidwingira,Nago,Anwar,Jin,Kaneyasu,Etter,Ying,Saitoh,Kashiwaya2019,Liu,Olthof,Jinepl,Asano2003}. For example, the experimental measurements of the critical current in the Au$_{0.5}$In$_{0.5}$/Sr$_2$RuO$_4$ junction indicate the odd-parity pairing symmetry in Sr$_2$RuO$_4$\cite{Nelson}. The subsequent theoretical calculations  suggest that the chiral $p$-wave pairing or the chiral singlet pairing in Sr$_2$RuO$_4$ can well explain the experimental finding\cite{Zutic,Asano2005}. The experimental observation of the differences in the magnetic field modulation of the critical current in the Ru/Sr$_2$RuO$_4$ junction confirms the $p$-wave pairing and its time-reversal symmetry breaking in Sr$_2$RuO$_4$\cite{Kidwingira}. A large change in supercurrent path also suggests the chiral $p$-wave state in the Ru/Sr$_2$RuO$_4$ interface\cite{Nago} while another experiment on the topological junction demonstrates multicomponent order parameters in the bulk Sr$_2$RuO$_4$\cite{Anwar}. The anomalous temperature dependence of the critical current and the proximity effect in the Ru/Sr$_2$RuO$_4$ junctions are also investigated experimentally and theoretically under the possible pairing symmetries\cite{Jin,Kaneyasu,Etter,Ying}. In addition, the inversion invariant behavior of the critical current under the reduced dimension is clarified experimentally in the Nb/Sr$_2$RuO$_4$ junction\cite{Saitoh} and the theoretical study reveals the helical $p$-wave symmetry of Sr$_2$RuO$_4$\cite{Kashiwaya2019}.

Besides the orbital part, the pairing wave function in TS also has the spin structure characterized by the $\bf{d}$-vector\cite{Balian}. The SS/TS Josephson junctions can provide the information of the spin structure. For example, the spontaneous spin accumulation due to the mismatch of the spin pairing symmetries in SS/TS points along the $\bf{d}$-vector in TS\cite{Sengupta}. The spin current and the spin accumulation in the more general SS/TS junctions with the $\bf{k}$-dependent $\bf{d}$-vector are also examined\cite{Lu}. The presence or absence of the Rabi oscillation in the supercurrent of the SS/TS junction dependes on the spin state of the junction\cite{Elster}. The control of the magnetism in the SS/TS junctions with TS being of the equal spin helical and the opposite spin chiral states are considered\cite{Romano}. The introduction of ferromagnetism in the SS/TS junctions offers another method to probe both the orbit part and the spin structure of the pairing wave function in TS\cite{Millis,Tanaka1,Tanaka2,Sothmann,Yokoyama,Brydon}. The presence of ferromagnetism in the SS/ferromagnet/TS junctions can bring rich CPRs, especially the lowest order current absent in the pure SS/TS junction due to the orthogonality of the spin pairing states in SS and TS\cite{Pals,Millis}. The types of the lowest order current strongly depend on the pairing symmetry of TS and the relative orientation between magnetization and the $\bf{d}$-vector. Furthermore, the $0$-$\pi$ transition and the $\phi_{0}$ phase can be realized in the SS/ferromagnet/TS junctions by tuning the magnitude of the exchange field or changing the direction of magnetization through an external field\cite{Yokoyama,Brydon}, which can be used for the design of the phase-based qubits\cite{Ioffe,Yamashita,Buzdin,Szombati,Cheng2019,Cheng2021}.

Nonetheless, there are still two questions to be solved in the detection of the pairing wave function of TS and the applications of the novel phases. The first is that the symmetry of the order parameter in TS continues to be debated\cite{Mackenzie}. The existing studies in the junctions containing SS and TS can not identify the specific form of the pairing wave function and even draw contradictory conclusions. Actually, various of technical means are recently employed to exert constraints on the identification of the order parameter in TS\cite{Mueller,Pustogow,Khasanov,Petsch,Grinenko,Hassinger,Jerzembeck,Ghosh,Benhabib,Grinenkonc,Grinenkoprb}. More effective platforms are still urgently needed to provide the more convincing evidence for the pairing wave function. The second is the inevitable influence of an applied field or ferromagnetism on the intrinsic pairing wave function in TS since the direction of the $\bf{d}$-vector can be easily rotated by a small magnetic field\cite{Annett}. In the SS/TS junctions, the detection of the spin-triplet superconductivity needs the magnetic flux produced by an external field\cite{Nelson,Kidwingira,Nago,Jin,Kaneyasu,Etter,Saitoh,Kashiwaya2019,Liu}. In the SS/ferromagnet/TS junctions\cite{Millis,Tanaka1,Tanaka2,Sothmann,Yokoyama,Brydon}, the stray field produced by ferromagnet cannot be fully eliminated\cite{Smejkal2}, which will not only change the $\bf{d}$-vector but also be destructive to the design of the qubits based on the novel phases. Moreover, there are reasons to believe that the existence of the second question impedes the resolution of the first question. How to resolve the two questions at once is an urgent issue in the condensed matter physics. Altermagnetic crystals possesses the vanishing magnetization and stray fields but still break the time-reversal symmetry as ferromagent does\cite{Smejkal2}, which pave the way for the identification of the intrinsic order parameter in TS and provide possibilities for the design of the field-free qubits based on the novel phases in the junctions containing SS and TS.

Here, we propose the SS/AM/TS Josephson junctions by
taking altermagnetism instead of ferromagnetism as the spin active mechanism. The interplay between altermagnetism and the unconventional spin-triplet superconductivity is systematically studied. Our junctions possess the zero net macroscopic magnetization and zero stray field, which are superior to the SS/ferromagnet/TS junctions with the inevitable competition between magnetization and superconductivity and the inevitable influence on the $\bf{d}$-vector by the stray field. However, the lowest order current in the SS/ferromagnet/TS junctions can still be realized in our junctions. Its type is also strongly dependent on the pairing wave functions in TS, which include valid information of the intrinsic order parameter in TS. Especially, the novel Josephson phase and the $0$-$\pi$ transition achieved in the SS/ferromagnet/TS junctions by tuning magnetization through an external field\cite{Yokoyama,Brydon} can also be realized in our junctions by adjusting the orientation of AM without any external field. Our junctions provide the feasible scheme to settle the two mentioned questions simultaneously.

In this paper, we focus on the orientation-dependent CPRs in the junctions.
The CPRs of the SS/AM/TS junctions are strongly dependent
on the crystallographic orientation of AM and
the types of the Cooper pairings.
The orientation of AM can be denoted by the angle $\alpha$
between the crystalline axis and the interface normal [see Fig.1(a)].
For the pure $d_{xy}$-wave altermagnetism with $\alpha =(2n+1)\pi/4$
with the integer number $n$, the CPRs of the SS/AM/TS junctions
are of the $\sin{2\phi}$ type.
In this case, the lowest order current is absent, which is irrespective of
the specific form of the Cooper pairing wave functions in SS and TS.
For the altermagnetism with $\alpha \not= (2n+1)\pi/4$,
the $\cos{\phi}$ type current can emerge for the $s$-wave ($d_{x^2-y^2}$-wave)
SS/AM/chiral $p$-wave ($p_x$-wave) TS junctions and
the $d_{xy}$-wave SS/AM/chiral $p_y$-wave TS junctions, as shown in Tab.I.
These junctions can host the $\phi_{0}$ phase with finite
current at the zero phase difference.
On the other hand, the $\sin{\phi}$ type current can appear
for the $d_{xy}$-wave SS/AM/chiral $p$-wave TS junctions,
and the $0$-$\pi$ transition can be obtained in the junctions
by adjusting the orientation angle $\alpha$.
In addition, the symmetric relations of the CPRs are obtained,
which are related with the variances of the Hamiltonians of SS, AM and TS
under the mirror reflection, the time-reversal and the spin rotation operations. These obtained results are peculiar to the interaction between AM and TS in the SS/AM/TS junctions, which are helpful in the detection of the spin-triplet superconductivity.

The rest of the paper is organized as follows. In Sec.$\text{\uppercase\expandafter{\romannumeral2}}$, we give the model of our junctions and the expression of the Josephson current. Sec.$\text{\uppercase\expandafter{\romannumeral3}}$ presents the numerical results of CPRs for different orientation angles in AM and different paring wave functions in SS and TS. Sec.$\text{\uppercase\expandafter{\romannumeral4}}$ gives the selection rules for the lowest order current and the symmetry analyses of CPRs. Sec.$\text{\uppercase\expandafter{\romannumeral5}}$ concludes this paper. The detailed expressions for the discretization of the continuum Hamiltonians are
relegated to the Appendix.

\section{\label{sec2}Model and Formulation}

\begin{figure}[!htb]
\centerline{\includegraphics[width=1\columnwidth]{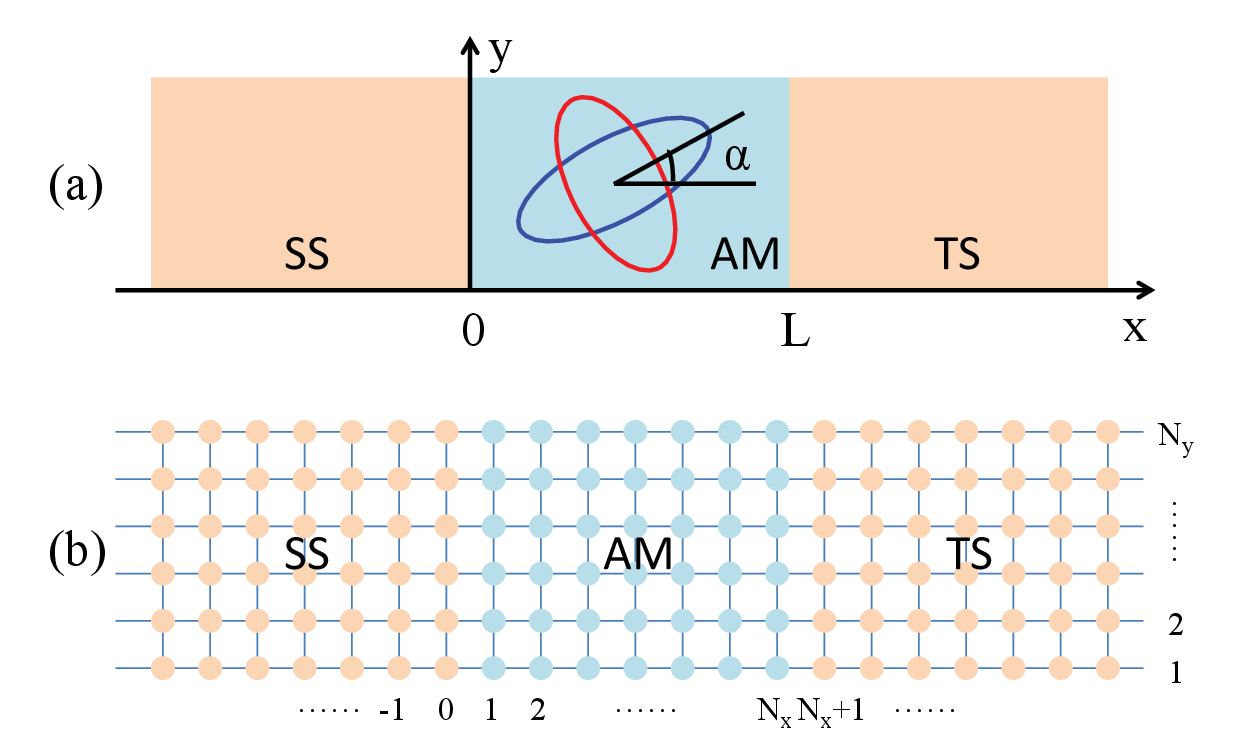}}
\caption{(a) The schematic illustration of the SS/AM/TS junctions.
The two ellipses in the AM region denote the Fermi surfaces for different spin of electrons. The blue one is for the down spin and the red one is for the up spin. The orientation of AM is described by the angle $\alpha$ between the crystalline axis (the major axis of the blue ellipse) and the normal direction of the interface. (b) The two-dimensional square lattice on which the SS/AM/TS junctions are discretized.}
\label{fig1}
\end{figure}

The SS/AM/TS Josephson junctions we consider are schematically shown
in Fig.1(a). The Josephson current flows parallel to the $x$ axis.
The SS/AM and AM/TS interfaces parallel to the $y$ axis
are located at $x=0$ and $x=L$.
The semi-infinite SS and TS are placed in the left region with $x<0$
and the right region with $x>L$, respectively.
For SS (TS), the Hamiltonian for the continuum model can be written as\cite{Tanaka74,Kashiwaya,Cheng103}
\begin{eqnarray}
H_{SS(TS)}=\sum_{{\bf{k}}}\Psi_{l(r){\bf{k}}}^{+}\check{H}_{ss(ts)}({\bf{k}})\Psi_{l(r){\bf{k}}}.\label{cHS}
\end{eqnarray}
Here, the operator $\Psi_{l(r){\bf{k}}}=(c_{l(r),{\bf{k}}\uparrow},c_{l(r),{\bf{k}}\downarrow},c_{l(r),{-\bf{k}}\uparrow}^{+},c_{l(r),{-\bf{k}}\downarrow}^{+})^{T}$ with the subscript $l(r)$ denoting the left SS (the right TS). The Bogoliubov-de Gennes (BdG) Hamiltonian $\check{H}_{ss(ts)}({\bf{k}})$ is given by
\begin{eqnarray}
\check{H}_{ss(ts)}({\bf{k}})=
\left(\begin{array}{cc}
\hat{h}(\bf{k})&\hat{\Delta}_{ss(ts)}(\bf{k})\\
-\hat{\Delta}_{ss(ts)}^{*}(-\bf{k})&-\hat{h}^{*}(-\bf{k})
\end{array}\right)
\end{eqnarray}
in the particle-hole$\otimes$spin space.
Here, $\hat{h}({\bf{k}})=[t_{0}(k_{x}^2+k_{y}^2)-\mu]\sigma_{0}$ with the two-dimensional wave vector ${\bf{k}}=(k_{x},k_{y})$ and the chemical potential $\mu$. For SS, the energy gap matrix $\hat{\Delta}_{ss}({\bf{k}})=\Delta_{0}f_{ss}({\bf{k}})e^{i\phi_{l}}i\sigma_{y}$ with $f_{ss}({\bf{k}})=1$ for the s-wave pairing, $f_{ss}({\bf{k}})=(k_{x}^2-k_{y}^2)$ for the $d_{x^2-y^2}$-wave pairing and $f_{ss}({\bf{k}})=2k_{x}k_{y}$ for the $d_{xy}$-wave pairing. For TS, the energy gap matrix $\Delta({\bf{k}})=(\vec{\sigma}\cdot{\bf{d}({\bf{k}})})i\sigma_y$ with ${\bf{d}}({\bf{k}})=\Delta_{0}e^{i\phi_{r}}f_{ts}({\bf{k}})\hat{z}$. Here, the ${\bf{d}}$ vector in TS has been assumed to be along the out-plane direction which is denoted by the unit vector $\hat{z}$. The orbit part $f_{ts}({\bf{k}})=\eta_{1}k_{x}+i\eta_{2}k_{y}$ with $(\eta_{1},\eta_{2})=(1,1)$ for the chiral $p$-wave pairing, $(\eta_{1},\eta_{2})=(1,0)$ for the $p_{x}$-wave pairing and $(\eta_{1},\eta_{2})=(0,-i)$ for the $p_{y}$-wave pairing. We have used $\Delta_{0}$ to denote the magnitude of the superconducting gap and $\phi_{l(r)}$ to denote the phase of the left SS (the right TS). $\sigma_{0}$ and ${\vec{\sigma}}=(\sigma_{x},\sigma_{y},\sigma_{z})$ are the identity matrix and the Pauli matrices in the spin space. We have taken $\hbar=1$ and the same gap magnitude in SS and TS for simplicity.

For AM in the central region with $0<x<L$, the Hamiltonian can be written as\cite{Smejkal2,Sun,Ouassou,Beenakker}
\begin{eqnarray}
H_{AM}=\sum_{\bf{k}}\psi({\bf{k}})^{+}\check{H}_{am}({\bf{k}})\psi({\bf{k}}),\label{cHA}
\end{eqnarray}
with the operator $\psi({\bf{k}})=(c_{{\bf{k}}\uparrow},c_{{\bf{k}}\downarrow},c_{{-\bf{k}}\uparrow}^{+},c_{{-\bf{k}}\downarrow}^{+})^{T}$ and the BdG Hamiltonian
\begin{eqnarray}
\check{H}_{am}({\bf{k}})=\left(\begin{array}{cc}
\hat{h}_{am}({\bf{k}})&0\\
0&-\hat{h}^{*}_{am}(-\bf{k})
\end{array}\right)
\end{eqnarray}
in the particle-hole$\otimes$spin space.
Here, $\hat{h}_{am}({\bf{k}})=\hat{h}({\bf{k}})+t_{J}[(k_{x}^2-k_{y}^2)\cos{2\alpha}+2k_{x}k_{y}\sin{2\alpha}]\sigma_{z}$ with the angle $\alpha$ between the crystalline axis and the SS/AM (AM/TS) interface normal.
The angle $\alpha$ defines the orientation of AM as shown in Fig.1(a).
For $\alpha=0$, AM is of the $d_{x^2-y^2}$-wave altermagnetism while for $\alpha=\pi/4$, AM is of the $d_{xy}$-wave altermagnetism\cite{Smejkal2}.

We discretize the above continuum Hamiltonians on a two-dimensional square lattice as shown in Fig.1(b). The lattice constant is taken as $a$. The discrete Hamiltonian is given by\cite{Cheng2021}
\begin{equation}
\begin{aligned}
H=\sum_{{\bf{i}}}[\Psi_{{\bf i}}^{+}\check{H}_{0}\Psi_{{\bf i}}
+\Psi_{{\bf i}}^{+}\check{H}_{x}\Psi_{{\bf i}+\delta x}
+\Psi_{{\bf i}}^{+}\check{H}_{y}\Psi_{{\bf i}+\delta y}\\
+\Psi_{{\bf i}}^{+}\check{H}_{xy}\Psi_{{\bf i}+\delta x+\delta y}
+\Psi_{{\bf i}}^{+}\check{H}_{x\bar{y}}\Psi_{{\bf i}+\delta x-\delta y}+H.c.].\label{dH}
\end{aligned}
\end{equation}
Here, ${\bf i}=(i_{x},i_{y})$ denotes the position of sites in the lattice with $i_{y}$ being limited in $1\le i_{y}\le N_{y}$ as shown in Fig.1(b).
The width $W$ of the junctions satisfies $W=(N_{y}-1)a$.
For SS (TS), one has $i_{x}\le 0$ ($i_{x}\ge N_{x}+1$) and the operator $\Psi_{{\bf i}}=(\Psi_{l(r){\bf i}\uparrow},\Psi_{l(r){\bf i}\downarrow},\Psi_{l(r){\bf i}\uparrow}^{+},\Psi_{l(r){\bf i}\downarrow}^{+})^T$ in SS (TS).
For AM, one has $1\le i_{x}\le N_{x}$ and the operator $\Psi_{{\bf i}}=(\psi_{{\bf i}\uparrow},\psi_{{\bf i}\downarrow},\psi_{{\bf i}\uparrow}^{+},\psi_{{\bf i}\downarrow}^{+})^{T}$.
The length $L$ of AM (the central region) satisfies $L=(N_{x}-1)a$.
The subscripts ${\bf i}+\delta x$ and ${\bf i}+\delta y$ represent the nearest neighbor sites of the ${\bf i}$th site along the $x$ direction and the $y$ direction, respectively.
The subscripts ${\bf i}+\delta x+\delta y$ and ${\bf i}+\delta x-\delta y$ represent the next nearest neighbor sites of the ${\bf i}$th site. The explicit expressions of $\check{H}_{0}$, $\check{H}_{x}$, $\check{H}_{y}$, $\check{H}_{xy}$ and $\check{H}_{x\bar{y}}$ for SS, TS and AM are presented in Appendix.

The hopping between different regions can be described by the following tunneling Hamiltonian
\begin{eqnarray}
H_{T} &= &\sum_{1\le i_{y}\le N_{y}}\left[\Psi_{(0,i_{y})}^{+}\check{T}\psi_{(1,i_{y})}\right.\nonumber\\
 & & +\left.\Psi_{(N_{x}+1,i_{y})}^{+}\check{T}\psi_{(N_{x},i_{y})}
+H.c.\right],\label{TH}
\end{eqnarray}
with the hoping matrix $\check{T}=\text{diag}(t,t,-t^*,-t^*)$.

The particle number operator for SS can be defined as
\begin{eqnarray}
N=\sum_{i_{x}\le 0}\sum_{\sigma=\uparrow,\downarrow}\Psi_{{\bf i}\sigma}^{+}\Psi_{{\bf i}\sigma}.
\end{eqnarray}
Using the Green's function method, the Josephson current can be given by\cite{Cheng2021,Sun2009,Li2018,Song2016}
\begin{eqnarray}
I=e\langle\frac{dN}{dt} \rangle=-\frac{e}{2\pi}\sum_{1\le i_{y}\le N_{y}}\int dE\text{Tr}[\Gamma_{z}\check{T}G^{<}(E,i_{y})+H.c.],\label{Jc}
\end{eqnarray}
with $\Gamma_{z}=\sigma_{z}\otimes\text{1}_{2\times2}$. The lesser Green's function $G^{<}(E,i_{y})$ is the Fourier transform of $G^{<}(t,t;(1,i_{y}),(0,i_{y}))$ which is defined as $G^{<}(t,t';(1,i_{y}),(0,i_{y}^{'}))=i\langle \Psi_{l(0,i_{y}^{'})}^{+}(t')\otimes\psi_{(1,i_{y})}(t)\rangle$. Due to the absence of the bias voltage on the SS/AM/TS junctions, the structure is in the equilibrium. From the fluctuation-dissipation theorem\cite{Cheng107}, we obtain
\begin{eqnarray}
G^{<}(E,i_{y})=-f(E)[G^{r}(E,i_{y})-G^{a}(E,i_{y})],\label{lG}
\end{eqnarray}
with the Fermi distribution function $f(E)$. The retard Green's function $G^{r}(E,i_{y})$ can be expressed by the retard Green's functions $G_{AM}^{r}(E,i_{y},j_{y})$ in AM and the free retard Green's function $g_{SS}^{r}(E,j_{y},i_{y})$ in SS\cite{Cheng2021}.
The latter two Green's functions can be obtained using $\check{H}_{0}$, $\check{H}_{x}$, $\check{H}_{y}$, $\check{H}_{xy}$ and $\check{H}_{x\bar{y}}$ in Eq.(\ref{dH}).

\section{\label{sec3}Numerical results for CPRs}
In the following numerical calculations, we have taken $t_{0}=1$ and $t_{J}=0.5$, which can well describe the anisotropy of altermagnetism.
The hopping magnitude $t$ between the SS (TS) and AM is taken as $t=t_0=1$.
For the chemical potential, we take the same value
in the three regions, i.e., $\mu=2$.
For the lattice parameters, $a=1$, $N_{x}=40$ and $N_{y}=40$ are considered. The phase difference is defined as $\phi=\phi_{l}-\phi_{r}$.
The unit of the Josephson current is taken as $e\Delta_0/2\pi$. We will take $\Delta_{0}=0.01$ and the coherence length $\xi=\hbar v_{F}/\Delta_0$ in superconductors can be calculated as about $283a$. Since the length of AM is given by $L=(N_x-1)a\ll\xi$, the short junctions are considered by us in this paper.

Before presenting the numerical results, we give the first universal symmetric relation satisfied by all junctions in this paper, i.e., $I(\alpha,\phi)=I(\pi+\alpha,\phi)$, which can be directly derived from the periodicity of altermagnetism presented in Eq.(\ref{cHA}). If one keeps SS and TS unchanged and only rotates AM by $\pi$ about $z$ axis, the junctions do not change. The Josephson current keeps invariant, which gives the universal symmetric relation. Next, we show the numerical results and discussions for CPRs according to the pairing wave functions in SS. The results for the $s$-wave SS are shown in Sec.$\text{\uppercase\expandafter{\romannumeral3}}A$ and those for the $d$-wave SS are shown in Sec.$\text{\uppercase\expandafter{\romannumeral3}}B$. For the both situations, three types of pairing wave functions in TS are considered, which are  the chiral $p$-wave, the $p_x$-wave and the $p_y$-wave states.

\subsection{\label{sec3.1}$s$-wave SS/AM/TS junctions}
\begin{figure}[!htb]
\centerline{\includegraphics[width=1\columnwidth]{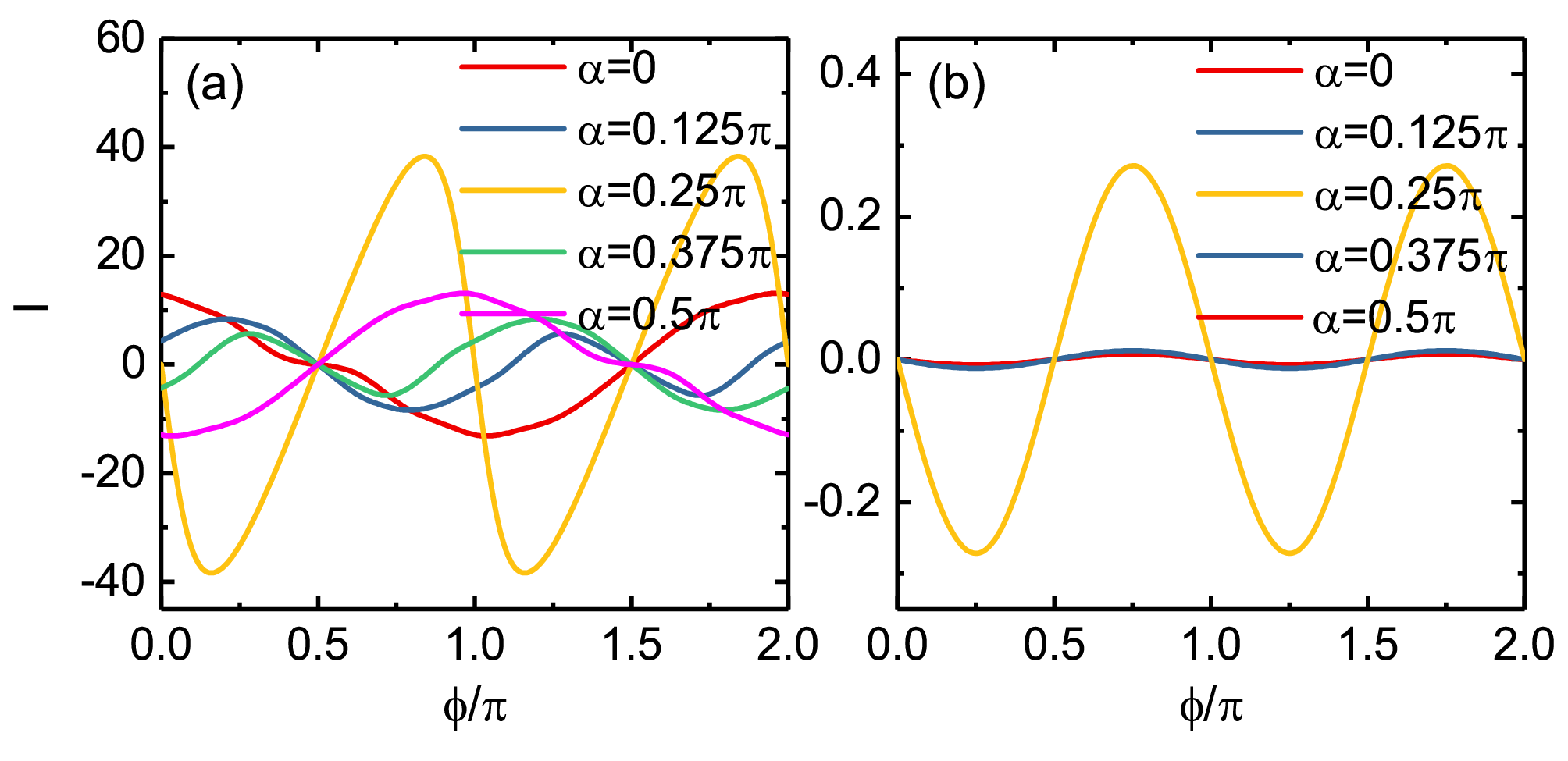}}
\caption{The CPRs for (a) the $s$-wave SS/AM/chiral $p$-wave TS junctions and
(b) the $s$-wave SS/AM/$p_y$-wave TS junctions.
In (b), the Josephson currents obey the relation $I(\alpha)=I(\pi/2-\alpha)$.}
\label{fig2}
\end{figure}

First, we present the results for the $s$-wave SS/AM/TS junctions.
Fig.2(a) shows the orientation-dependent CPRs for the chiral $p$-wave TS.
The CPRs strongly depend on the orientation angle $\alpha$ of AM.
Generally, the Josephson current can be expressed as the sum of the Fourier series, i.e., $I(\phi)=\sum_{n\ge1}[a_{n}\sin{n\phi}+b_{n}\cos{n\phi}]$, where $a_{n}$ and $b_{n}$ are the expansion coefficients and $n$ is an integer number.
For $\alpha=\pi/4$, the CPR is of the $\sin{2\phi}$ type
and the lowest order current $\sin{\phi}$ and $\cos{\phi}$
are absent as shown in Fig.2(a).
However, when the orientation angle $\alpha$ deviates from $\alpha=\pi/4$,
the $\cos{\phi}$ type current will happen
and the current is nonzero at the phase difference $\phi=0$ [see Fig.2(a)].
This means that the $\phi_{0}$ phase is formed in
the $s$-wave SS/AM/chiral $p$-wave TS junctions.
Since the current is still vanishing at $\phi=\pi/2$ and
$\phi=3\pi/2$ [see Fig.2(a)],
the Josephson current does not include the $\sin{\phi}$ term.
This indicates that the junction can hold a $\phi_0$ phase
with $\phi_0=\pi/2$ or $3\pi/2$.
We don't give the numerical results for
the lager orientation angle $\alpha$ of AM
because the CPRs for the $s$-wave SS/AM/chiral $p$-wave TS junction
satisfy the following symmetric relations,
\begin{eqnarray}
I(\alpha,\phi)&=&I(\pi/2+\alpha,\pi+\phi),\label{safcs1}\\
I(\alpha,\phi)&=&-I(\pi-\alpha,\pi-\phi).\label{safcs2}
\end{eqnarray}
The combination of Eqs.(\ref{safcs1}) and (\ref{safcs2}) leads to another relation $I(\alpha,\phi)=-I(\pi/2-\alpha,2\pi-\phi)$, which is demonstrated by the CPRs in Fig.2(a).
In addition, from Fig.2(a), one can also see that the CPRs satisfy the relation
\begin{eqnarray}
I(\alpha,\phi)&=&-I(\alpha,\pi-\phi).\label{safcs3}
\end{eqnarray}
This symmetric relation makes that the forward and reverse supercurrents
have the same behavior although the junction is left-right inversion asymmetric.
These symmetric relations of CPRs in Eqs.(\ref{safcs1}-\ref{safcs3}) are closely related with the transformations of the continuum Hamiltonians for SS, AM and TS under some operations, which will be discussed in details in Sec.$\text{\uppercase\expandafter{\romannumeral4}}$.

For the $s$-wave SS/AM/$p_{x}$-wave TS junctions, the characters and the symmetric relations of CPRs are same with those in Fig.2(a) for the $s$-wave SS/AM/chiral $p$-wave TS junctions, which are not presented here.

Fig.2(b) shows the CPRs for the $s$-wave SS/AM/$p_y$-wave TS junctions.
In this situation, all CPRs for different orientation angles $\alpha$
are of the $\sin{2\phi}$ type although the orientation of AM can change the critical value of the Josephson current.
The CPRs for the $p_{y}$-wave TS junctions satisfy the following relations
\begin{eqnarray}
I(\alpha,\phi)&=&I(\pi/2+\alpha,\pi+\phi),\label{safpys1}\\
I(\alpha,\phi)&=&-I(\pi-\alpha,2\pi-\phi),\label{safpys2}\\
I(\alpha,\phi)&  =& -I(\alpha,\pi-\phi) = I(\alpha,\pi+\phi).\label{safpys3}
\end{eqnarray}
The combination of Eqs.(\ref{safpys1}) and (\ref{safpys2}) brings another
relation which is $I(\alpha,\phi)=-I(\pi/2-\alpha,\pi-\phi)$.
Then by furthermore combining with Eq.(\ref{safpys3}),
we have $I(\alpha,\phi)=I(\pi/2-\alpha,\phi)$ for the $p_y$-wave TS junctions,
which can be seen from the CPRs in Fig.2(b).

\subsection{\label{sec3.1}$d$-wave SS/AM/TS junctions}

\begin{figure}[!htb]
\centerline{\includegraphics[width=1\columnwidth]{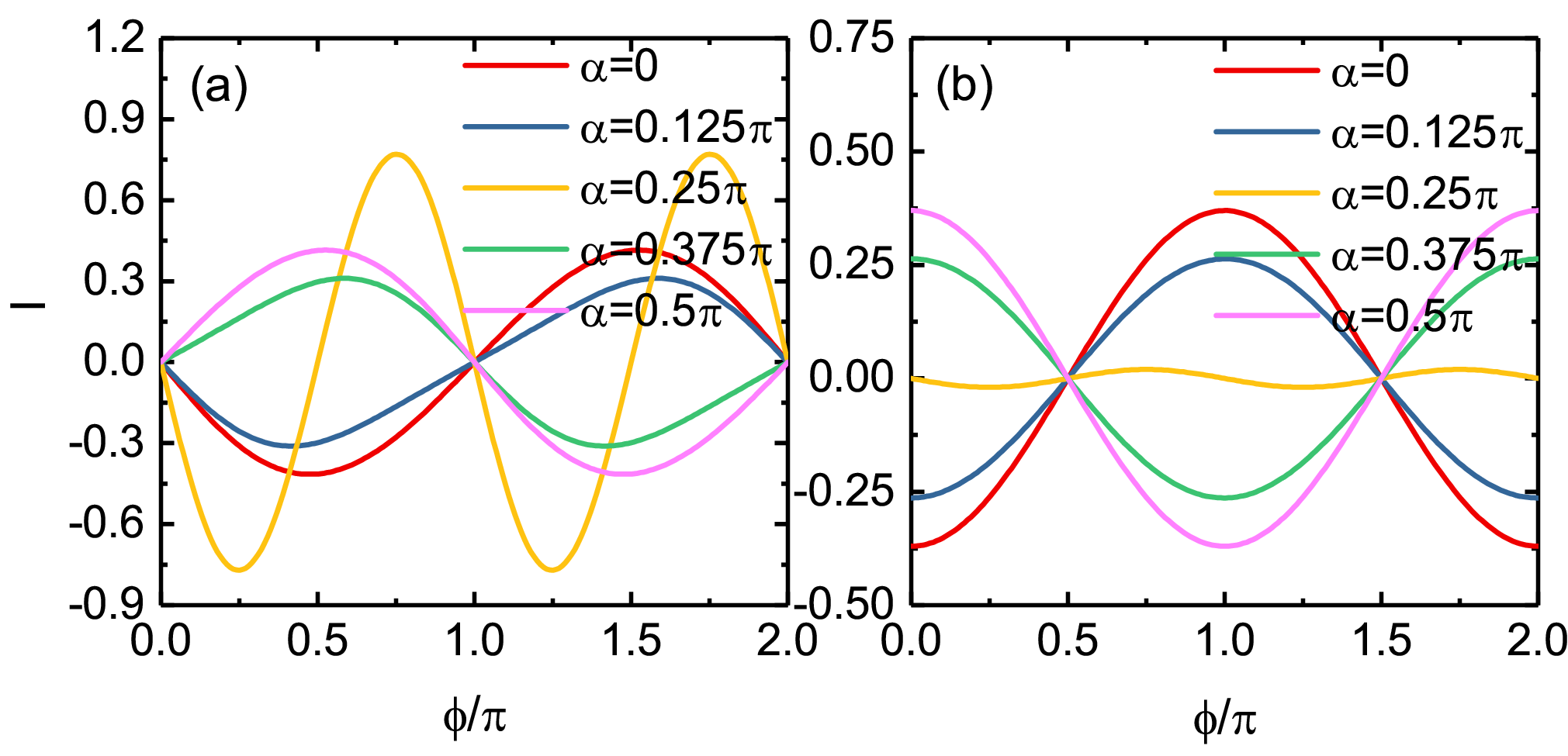}}
\caption{The CPRs for (a) the $d_{xy}$-wave SS/AM/chiral $p$-wave TS junctions
and (b) the $d_{xy}$-wave SS/AM/$p_y$-wave TS junctions.}
\label{fig3}
\end{figure}

In this subsection, we focus on the $d_{xy}$-wave SS/AM/TS junctions.
Fig.3(a) shows the CPRs of the $d_{xy}$-wave SS/AM/chiral $p$-wave TS junctions
for different orientation angles $\alpha$ of AM.
For $\alpha=\pi/4$, the CPR is of the $\sin{2\phi}$ type
with the absence of the lowest order currents $\sin{\phi}$ and $\cos{\phi}$,
which is the same with the $s$-wave SS case as shown in Fig.2.
But, distinct from the $s$-wave SS situation,
the deviation of the orientation angle from $\alpha=\pi/4$
for the $d_{xy}$-wave SS causes the emergence of the $\sin{\phi}$
type current as shown in Fig.3(a).
The zero current at $\phi=0$ indicates that the $\cos{\phi}$
type current is impossible to exist.
Furthermore, the current at the interval $\phi\in (0,\pi)$ for
$\alpha>\pi/4$ is positive, while it is negative for $\alpha<\pi/4$.
The former corresponds to the $0$ phase of the junctions
and the latter corresponds to the $\pi$ phase of the junctions.
In other words, the change of the orientation angle $\alpha$ of AM
can realize the $0$-$\pi$ transition in the
$d_{xy}$-wave SS/AM/chiral $p$-wave TS junctions.
The CPRs for the $d_{xy}$-wave SS/AM/chiral $p$-wave TS junctions
respect the symmetric relations in Eqs.(\ref{safpys1}) and (\ref{safpys2})
for the $s$-wave SS/AM/$p_y$-wave TS junctions. In addition, it has
the relation $I(\alpha,\phi)=-I(\alpha,2\pi-\phi)$ [see Fig.3(a)],
as in the conventional SS/normal conductor/SS junction\cite{Golubov}.

For the $d_{xy}$-wave SS/AM/$p_{x}$-wave TS junctions,
the lowest order current does not exist
and the CPRs are only of the $\sin{2\phi}$ form similar
to the results for the $s$-wave SS/AM/$p_y$-wave TS junctions in Fig.2(b).
Furthermore, the CPRs also obey the symmetric relations given in Eqs.(\ref{safpys1}) and (\ref{safpys2}).
For these reasons, we don't show the numerical results for
the $d_{xy}$-wave SS/AM/$p_{x}$-wave TS junctions.

The CPRs for the $d_{xy}$-wave SS/AM/$p_{y}$-wave TS are shown in Fig.3(b). Different from the results for the chiral $p$-wave TS in Fig.3(a),
the $\cos{\phi}$ type current appears in the $p_{y}$-wave TS junctions
when $\alpha\ne\pi/4$.
There is no $\sin{\phi}$ type current since the current
is zero at $\phi=\pi/2$ and $\phi=3\pi/2$.
This indicates that the $\phi_{0}$ phase with $\phi_{0}=\pi/2$ or $3\pi/2$
can be formed in the $d_{xy}$-wave SS/AM/$p_y$-wave TS junctions.
The CPRs for the junctions obey the symmetric relations in Eqs.(\ref{safcs1})-(\ref{safcs3}) for the $s$-wave SS/AM/chiral
$p$-wave TS junctions.

As for the $d_{x^2-y^2}$-wave SS/AM/TS junctions,
the CPRs exhibit the similarities with those for the $s$-wave SS/AM/TS junctions.
The $\cos{\phi}$ type current can happen for the chiral $p$-wave TS and the $p_x$-wave TS except for $\alpha=\pi/4$, as in Fig.2(a).
The $\phi_{0}$ phase can be realized in these junctions.
For the $d_{x^2-y^2}$-wave SS/AM/$p_y$-wave TS junctions,
the Josephson current is of the $\sin{2\phi}$ type,
as in $s$-wave SS/AM/$p_y$-wave TS junctions [see Fig.2(b)].
The junctions for the $d_{x^2-y^2}$-wave SS respect the same symmetric relations with the $s$-wave SS, which are presented in Eqs.(\ref{safcs1})-(\ref{safpys3}).
Finally, we can get the second universal relation $I(\alpha,\phi)=I(\pi/2+\alpha,\pi+\phi)$ presented in Eqs.(\ref{safcs1}) and (\ref{safpys1}), which is satisfied by all SS/AM/TS junctions in this paper.

\section{\label{sec4}Selection rules and symmetry analyses}
\subsection{\label{sec4.1}Selection rules}

\begin{table}[!htb]
\caption{The selection rules for the lowest order current in the SS/AM/TS junctions. The symbol ``$-$" means the absence of the corresponding lowest order current and $n$ is an integer number.}
\label{table.1}
\begin{center}
\renewcommand{\multirowsetup}{}
\begin{tabular}{cccc}
\hline\hline
~~~~~SS~~~~~ & ~~~~~TS~~~~~ & ~~~~~$\cos{\phi}$~~~~~ & ~~~~~$\sin{\phi}$~~~~~ \\
\hline
$s$, $d_{x^2-y^2}$ & $p_{x}+ip_{y}$,$p_{x}$ & $\alpha\ne(2n+1)\frac{\pi}{4}$ & -- \\
$s$, $d_{x^2-y^2}$ & $p_{y}$ & -- & -- \\
$d_{xy}$ & $p_{x}+ip_{y}$ &--&$\alpha\ne(2n+1)\frac{\pi}{4}$  \\
$d_{xy}$ & $p_{y}$ & $\alpha\ne(2n+1)\frac{\pi}{4}$ & --\\
$d_{xy}$ & $p_{x}$ & -- & --\\
\hline\hline\label{tab1}
\end{tabular}
\end{center}
\end{table}

The lowest order current is the indispensable component
for the formation of the $\phi_{0}$ phase and the $0$-$\pi$ transition.
In Tab.{\ref{tab1}}, we summarize the selection rules for the $\sin{\phi}$
and $\cos{\phi}$ type currents in the SS/AM/TS junctions
based on the numerical results and the discussions in Secs.$\text{\uppercase\expandafter{\romannumeral3}}$.
It is found that $\alpha\ne\pi/4$ in AM is the necessary condition
for the emergence of the lowest order current besides
the specific pairing wave functions in SS and TS.
Taking the universal relations $I(\alpha,\phi)=I(\pi+\alpha,\phi)$ and $I(\alpha,\phi)=I(\pi/2+\alpha,\pi+\phi)$ into account, the condition can be generalized to $\alpha\ne(2n+1)\pi/4$ with $n$ being an integer number.
In other words, the orientation angle $\alpha$ of AM cannot
be equal to the odd multiples of $\pi/4$ in order to obtain
the lowest order current. On the other hand, the presence and absence of the lowest order current and its type are strongly dependent on the pairing wave functions in TS as shown in Tab.{\ref{tab1}}, which provide an effective method to detect the intrinsic order parameter of TS without the influences of the stray field and the external field.
Actually, the selection rules here in our junctions are very similar
to those for the SS/ferromagnet/TS junctions in Ref.[\onlinecite{Brydon}].
It is interesting to compare the mechanisms for the selection rules in the two kinds of junctions.

If we substitute a ferromagnet for AM in our junctions,
the necessary condition for the lowest order current will become
$\theta_{m}\ne\pi/2$ with $\theta_m$ being the polar angle
of the magnetization in ferromagnet.
Note that the polar angle $\theta_m$ here in our coordinate system
is just the azimuthal angle in the coordinate system of Ref.[\onlinecite{Brydon}].
In other words, the magnetization in ferromagnet must have
the component parallel to the out-plane direction
in order to obtain the lowest order current.
In this paper, the ${\bf{d}}$ vector in TS has been assumed
to be along the out-plane direction.
The in-plane magnetization is perpendicular to the $\bf{d}$ vector,
which is detrimental for the formation of the lowest order current.
This is because that TS will be of the equal-spin pairing state
with the in-plane spin equal to $\pm 1$ if the direction of
the in-plane magnetization is chosen as the spin quantization axis.
In this situation, the wave function of Cooper pairs
$|S=0,S_{\parallel}=0 \rangle$ in SS and
that $|S=1,S_{\parallel}=\pm 1 \rangle$ in TS are orthogonal
since SS is of the opposite spin pairing state
with the total spin equal to $0$. As a result,
the tunneling of a single Cooper pair from the left SS
to the right TS is forbidden.
The coherent tunneling of two Cooper pairs is permitted,
which only leads to the $\sin{2\phi}$ type current.
However, the out-plane magnetization parallel to the $\bf{d}$ vector
is helpful to form the lowest order current.
In this situation, TS will be of the opposite spin pairing state
$|S=1,S_{z}= 0 \rangle=\frac{\sqrt{2}}{2}\left(|S=1,S_{\parallel}= 1 \rangle
+|S=1,S_{\parallel}= -1 \rangle\right)$
with the out-plane spin equal to $0$.
The tunneling of a single Cooper pair $|S=0,S_{z}= 0 \rangle$
from the left SS to the right TS with the states $|S=1,S_{z}= 0 \rangle$
is possible.
Therefore, the rotation of the magnetization can tune the CPRs
through changing the out-plane component of the magnetization in ferromagnet.
This is the mechanism for the selection rules in the SS/ferromagnet/TS junctions.

In order to understand the mechanism for the SS/AM/TS junctions in this paper,
we can treat AM as a ferromagnetic system having an effective magnetization
${\bf{M}}_{\text{eff}}$ with its direction being fixed along the $z$ axis.
As shown in Eq.(\ref{cHA}), the effective magnetization
can be expressed as ${\bf{M}}_{\text{eff}}({\bf{k}})=t_{J}[(k_{x}^2-k_{y}^2)\cos{2\alpha}+2k_{x}k_{y}\sin{2\alpha}]\hat{z}$. The magnetization only has the out-plane component
parallel to the $\bf{d}$ vector.
However, this doesn't mean that the lowest order current
is always present in our junctions.
Different from the isotropic magnetization in ferromagnet,
the magnetization in AM has an anisotropic orbital part
which is $[(k_{x}^2-k_{y}^2)\cos{2\alpha}+2k_{x}k_{y}\sin{2\alpha}]$.
This difference resembles the distinction between
the $s$-wave SS and the $d$-wave SS.
The pairing in the former is isotropic
while the pairing in the latter is anisotropic.
The orbital part in AM provides a degree of freedom for AM
to tune the magnitude of the out-plane magnetization.
This degree of freedom is just the orientation angle $\alpha$ in AM.
For ease of understanding, we take the one-dimensional transport
with the incident angle $\theta=0$
as an example to discuss the effects of the orientation angle $\alpha$.
In this situation, one has $(k_x,k_y)/\sqrt{k_x^2+k_y^2}=(1,0)$
and the orbital part becomes $\cos{2\alpha}$.
For $\alpha=0$, the out-plane magnetization in AM reaches its maximum value
which corresponds to the polar angle $\theta_m=0$ of the ferromagnet case.
The appearance of the lowest order current is possible.
When one alters the orientation angle from $\alpha=0$ to $\alpha=\pi/4$,
the out-plane magnetization is decreased and finally disappears.
The lowest order current for $\alpha=\pi/4$ can not be expected.
The orientation angle $\alpha=\pi/4$ in AM corresponds to
the polar angle $\theta_{m}=\pi/2$ of the ferromagnet case.
However, $\theta_{m}=\pi/2$ does not change the ferromagnetic property
of ferromagnet, which just leads to the vanishing of the out-plane component
of magnetization not magnetization itself.
In contrast, $\alpha=\pi/4$ changes AM from the state
with the finite effective magnetization to the state
with the vanishing effective magnetization.
For $\alpha=\pi/4$, AM behaves like a normal metal without
any spin polarization since the orbital part $\cos{2\alpha}$ is equal to $0$.
In this case, the spin quantization axis can be selected arbitrarily.
If we chose the in-plane direction which is perpendicular to
the $\bf{d}$ vector in TS as the spin quantization axis,
TS will be of the equal-spin pairing state with the in-plane
spin equal to $\pm 1$.
The wave function of Cooper pairs $|S=0,S_{\parallel}=0 \rangle$ in SS
and that $|S=1,S_{\parallel}=\pm 1\rangle$ in TS becomes orthogonal.
The lowest order current is impossible to occur.
This is the mechanism for the selection rules in the SS/AM/TS junctions.

Now, we try to give a more comprehensive discussions about
the selection rules by constructing the free energy $\mathcal{F}$
of the SS/AM/TS junctions.
First, we introduce ${\bf{S}}({\bf{k}})=\Delta_{0}f_{ss}({\bf{k}})\Delta_{0}f_{ts}({\bf{k}})\hat{z}$, which includes the information of the interface spin
for the junctions between SS and TS.\cite{Brydon}
For the two dimensional transport, the incident electrons
possess different values of $k_y$.
We define the average of ${\bf{S}}({\bf{k}})$ as $\langle {\bf{S}}({\bf{k}})\rangle=\int_{-k_0}^{k_0}{\bf{S}}({\bf{k}})d{k_y}/\int_{-k_0}^{k_0} d{k_y}$.
%For the junctions being infinite along the $y$ direction, the integral interval $[-k_0,k_0]$ will become $[-\pi/2,\pi/2]$.
We can also define $\langle {\bf{M}}_{\text{eff}}({\bf{k}})\rangle$ as the average of the effective magnetization in AM in a similar way.
After a simple calculation, one can find $\langle {\bf{M}}_{\text{eff}}({\bf{k}})\rangle\varpropto\cos{2\alpha}\hat{z}$.
The interaction between the interface spin and the effective magnetization will contribute to the free energy and the lowest order current of the SS/AM/TS junctions. We speculate that the free energy possesses the following form\cite{Brydon}
\begin{eqnarray}
\begin{split}
\mathcal{F}\varpropto&\text{Re}[\langle {\bf{S}}({\bf{k}})\rangle\cdot\langle {\bf{M}}_{\text{eff}}({\bf{k}})\rangle] \sin{\phi}\\
+&\text{Im}[\langle {\bf{S}}({\bf{k}})\rangle\cdot\langle {\bf{M}}_{\text{eff}}({\bf{k}})\rangle] \cos{\phi}.\label{FE}
\end{split}
\end{eqnarray}
As the derivative of the free energy with respect to $\phi$,
the Josephson current can be accordingly written as
\begin{eqnarray}
\begin{split}
I\varpropto&\text{Re}[\langle {\bf{S}}({\bf{k}})\rangle\cdot\langle {\bf{M}}_{\text{eff}}({\bf{k}})\rangle] \cos{\phi}\\
+&\text{Im}[\langle {\bf{S}}({\bf{k}})\rangle\cdot\langle {\bf{M}}_{\text{eff}}({\bf{k}})\rangle] \sin{\phi}.
\end{split}
\end{eqnarray}
Taking $\langle {\bf{M}}_{\text{eff}}({\bf{k}})\rangle\varpropto\cos{2\alpha}\hat{z}$ into account, the Josephson current can be further expressed as $I\varpropto (\text{Re}[\langle {\bf{S}}({\bf{k}})\rangle]\cos{\phi}+\text{Im}[\langle {\bf{S}}({\bf{k}})\rangle]\sin{\phi})\cos{2\alpha}$. From this expression, we can find the lowest order current is vanishing when $\alpha=(2n+1)\pi/4$, which is independent on the paring wave functions of superconductors involved in $\langle {\bf{S}}({\bf{k}})\rangle$. For $\alpha\ne(2n+1)\pi/4$, the type of the lowest order current is determined by the specific pairing wave functions in SS and TS. For the $s$-wave and the $d_{x^2-y^2}$-wave SSs, if TS is of the chiral $p$-wave or $p_x$-wave pairing, $\langle {\bf{S}}({\bf{k}})\rangle$ is a real number and only the $\cos{\phi}$ type current can be obtained. If TS is of the $p_y$-wave pairing, $\langle {\bf{S}}({\bf{k}})\rangle=0$ and there is no the lowest order current in the junctions. For the $d_{xy}$-wave SS, if TS is of the chiral $p$-wave pairing, $\langle {\bf{S}}({\bf{k}})\rangle$ is a pure imaginary number and only the $\sin{\phi}$ type current can be obtained. If TS is of the $p_y$-wave pairing, $\langle {\bf{S}}({\bf{k}})\rangle$ is a real number and only the $\cos{\phi}$ type current can be obtained. If TS is of the $p_x$-wave pairing, $\langle {\bf{S}}({\bf{k}})\rangle=0$ and there is no the lowest order current in the junctions. These discussions are well consistent with the numerical results of CPRs in Fig.\ref{fig2} and Fig.\ref{fig3} and the selection rules summarized in Tab.\ref{tab1}. The free energy in Eq.(\ref{FE}) reflects the peculiar interplay between altermagnetism and the spin-triplet superconductivity, which provide the physical basis for the detection of the intrinsic order parameter of TS.

\subsection{\label{sec4.3}Symmetry analyses}

Now, we derive the symmetric relations from the mirror reflection operation about the $xz$ plane ${\mathcal{M}}_{xz}$, the time-reversal operation ${\mathcal{T}}$ and the spin rotation operation about the $y$ axis $\mathcal{R}_{y}(\pi)$. The effects of the operations on the annihilation operators $c_{{\bf{k}}s}$ with the spin $s$ can be given by\cite{Cheng2021,Cheng2017,yan1}
\begin{eqnarray}
&&\mathcal{M}_{xz}c_{(k_x,k_y)s}\mathcal{M}_{xz}^{-1}=sc_{(k_x,-k_y)\bar{s}},\label{Mo}\\
&&\mathcal{T}c_{{\bf{k}}s}\mathcal{T}^{-1}=sc_{-{\bf{k}}\bar{s}},\label{To}\\
&&\mathcal{R}_{y}(\pi)c_{{\bf{k}}s}\mathcal{R}_{y}(\pi)^{-1}=\bar{s}c_{{\bf{k}}\bar{s}}.\label{Ro}
\end{eqnarray}

Now, we consider the transformations of the Hamiltonian in each region under the above operations.
For the spin rotation about the $y$ axis, we have
\begin{eqnarray}
\mathcal{R}_{y}(\pi)H_{SS}(\phi_{l})\mathcal{R}_{y}(\pi)^{-1}=H_{SS}(\phi_{l}),\label{RSS}
\end{eqnarray}
for SS ,
\begin{eqnarray}
\mathcal{R}_{y}(\pi)H_{AM}(\alpha)\mathcal{R}_{y}(\pi)^{-1}=H_{AM}(\pi/2+\alpha),\label{RAF}
\end{eqnarray}
for AM and
\begin{eqnarray}
\mathcal{R}_{y}(\pi)H_{TS}(\phi_r)\mathcal{R}_{y}(\pi)^{-1}=H_{TS}(\pi+\phi_r),\label{RTS}
\end{eqnarray}
for TS. As a unitary operation, the spin rotation will not change the current. As a result, we can get $I(\alpha,\phi)=I(\pi/2+\alpha,\pi+\phi)$ from Eqs.(\ref{RSS})-(\ref{RTS}), which is the second universal symmetric relation as presented in Eqs.(\ref{safcs1}) and (\ref{safpys1}) satisfied by all junctions.

We define the joint operation $\mathcal{X}=\mathcal{T}\mathcal{M}_{xz}$ using the mirror reflection $\mathcal{M}_{xz}$ and the time-reversal operation $\mathcal{T}$ in Eqs. (\ref{Mo}) and (\ref{To}). Under the joint operation, we have
\begin{eqnarray}
\mathcal{X}H_{SS(TS)}(\phi_{l(r)})\mathcal{X}^{-1}=H_{SS(TS)}(-\phi_{l(r)}),
\end{eqnarray}
for the $s$-wave SS, the $d_{x^2-y^2}$-wave SS, and the $p_y$-wave TS,
while
\begin{eqnarray}
\mathcal{X}H_{SS(TS)}(\phi_{l(r)})\mathcal{X}^{-1}=H_{SS(TS)}(\pi-\phi_{l(r)}),
\end{eqnarray}
for the $d_{xy}$-wave SS, the chiral $p$-wave TS, and the $p_x$-wave TS.
The same operation can also gives
\begin{eqnarray}
\mathcal{X}H_{AM}(\alpha)\mathcal{X}^{-1}=H_{AM}(\pi-\alpha),
\end{eqnarray}
for AM. The mirror reflection will not change the current but the time reversal operation can inverse the current. As a result, we obtain the relation $I(\alpha,\phi)=-I(\pi-\alpha,\pi-\phi)$ for the $s$(or $d_{x^2-y^2}$)-wave SS/AM/chiral $p$-wave TS junctions, $s$(or $d_{x^2-y^2}$)-wave SS/AM/$p_x$-wave TS junctions and the $d_{xy}$-wave SS/AM/$p_y$-wave TS junctions while the relation $I(\alpha,\phi)=-I(\pi-\alpha,2\pi-\phi)$ holds for the $s$(or $d_{x^2-y^2}$)-wave SS/AM/$p_y$-wave TS junctions, the $d_{xy}$-wave SS/AM/chiral $p$-wave TS junctions and the $d_{xy}$-wave SS/AM/$p_x$-wave TS junctions. The two relations are just the ones presented in Eqs.(\ref{safcs2}) and (\ref{safpys2}), respectively.

Furthermore, we consider the joint operation $\mathcal{Y}=\mathcal{T}\mathcal{R}_{y}$. Under this joint operation,
the Hamiltonian $H_{AM}$ in the AM region is invariant,
but the Hamiltonians $H_{SS}$ and $H_{TS}$ will change
according to the following relations
\begin{eqnarray}
\mathcal{Y} H_{SS}(\phi_l)
\mathcal{Y}^{-1} &=& H_{SS}(-\phi_l)
\end{eqnarray}
for the $s$-wave SS, the $d_{x^2-y^2}$-wave SS, and the $d_{xy}$-wave SS;
and
\begin{eqnarray}
\mathcal{Y}H_{TS}(\phi_r)
\mathcal{Y}^{-1} &=& H_{TS}(\pi-\phi_r)
\end{eqnarray}
for the $p_x$-wave TS and the $p_y$-wave TS.
Therefore, we have the relation
$I(\alpha,\phi)=-I(\alpha,\pi-\phi)$
for the $s$-wave (or $d_{x^2-y^2}$-wave, $d_{xy}$-wave)
SS/AM/$p_x$-wave (or $p_y$-wave) TS junction,
as shown in Eq.(\ref{safpys3}) and in Figs. 2(b) and 3(b).

Finally, we give a short discussion about the experimental realization of our results. The orientation-dependent physical effects have been studied in many systems. One typical example is the transport in the $d$-wave superconductor junctions. For example, the tunneling conductance\cite{Tanaka1996}, the Josephson effects\cite{Tanaka1995,Tanaka1994} and the diode effect\cite{Tanaka2022} for the different crystallographic orientation angles of the $d$-wave superconductor
have been clarified in theory.
In these junctions, the continuous regulation of the orientation angle may be difficult in experiment. However, the transport quantities for several specific orientation angles can be measured experimentally. The conductances for different orientation angles of the $d$-wave superconductor have been detected in the normal metal/superconductor junctions\cite{Wang1999,Iguchi2000,Alff1998}.
The Josephson effect in the $s$-wave superconductor/insulator/$d$-wave superconductor junctions are also investigated for different well-defined geometries\cite{Chesca}.
In this paper, we have obtained the selection rules and the symmetric relations of CPRs in the SS/AM/TS junctions, which give the explicit values of the orientation angle for the $\pi$ phase and the $\phi_{0}$ phase.
It is believed that the two phases can be experimentally realized under the specific orientation angle.
On the other hand, the measurements of the Josephson current for several orientation angles are also enough to detect the intrinsic pairing wave function in TS.

\section{\label{sec4}Conclusions}

In contrast to ferromagnet, altermagnet provides a different degree of freedom which is its orientation.
For the SS/AM/TS junctions, the change of the orientation angle of AM
can activate the lowest order current and alter the CPRs in the junctions.
The $\phi_0$ phase and the $0$-$\pi$ transition can be realized,
which are the basic components for the design of field-free quantum devices.
The selection rules for the lowest order current are presented,
which give the necessary orientation angle of AM and the specific pairing wave functions in SS and TS for the formation of the lowest order current.
The mechanisms for the selection rules in the SS/AM/TS junctions
and the SS/feromagnet/TS junctions and their difference are clarified.
We also investigate the symmetric relations obeyed by the CPRs
in the SS/AM/TS junctions and their close connections with the mirror reflection, the time-reversal and the spin rotation operations.
Our results exhibit the peculiar transport signal
for the SS/AM/TS junctions and the novel interplay between altermagnetism and the spin-triplet superconductivity, which offer the practical feasibility of the detection of the Cooper pair wave function in TS.

\section*{\label{sec5}ACKNOWLEDGMENTS}

This work was financially supported
by the National Natural Science Foundation of China
under Grants Nos. 12374034, 11921005 and 11447175,
the Innovation Program for Quantum Science and Technology (2021ZD0302403),
the Strategic Priority Research Program of Chinese Academy of Sciences (XDB28000000)
and the project ZR2023MA005 supported by Shandong Provincial Natural Science Foundation. We acknowledge the Highperformance
Computing Platform of Peking University for
providing computational resources.

\section*{\label{sec5} Appendix}
\setcounter{equation}{0}
\renewcommand{\theequation}{A.\arabic{equation}}
The discrete Hamiltonian in Eq.(\ref{dH}) can be obtained from the continuum Hamiltonians in Eqs.(\ref{cHS}) and (\ref{cHA}) by discretizing them on the two-dimensional lattice. Now, we present the explicit expressions of $\check{H}_{0}$, $\check{H}_{x}$, $\check{H}_{y}$, $\check{H}_{xy}$ and $\check{H}_{x\bar{y}}$ in Eq.(\ref{dH}) for SS, TS and AM.

For the $s$-wave SS,
\begin{eqnarray}
\check{H}_{0}=\left(\begin{array}{cc}
(\frac{4t_{0}}{a^2}-\mu)\sigma_{0}&\Delta_{0} e^{i\phi_{l}}i\sigma_{y}\\
-\Delta_{0}e^{-i\phi_{l}}i\sigma_{y}&(-\frac{4t_{0}}{a^2}+\mu)\sigma_{0}
\end{array}\right),\label{sH0}
\end{eqnarray}
\begin{eqnarray}
\check{H}_{x}=\check{H}_{y}=\left(\begin{array}{cc}
-\frac{t_{0}}{a^2}\sigma_{0}&0\\
0&\frac{t_{0}}{a^2}\sigma_{0}
\end{array}\right),\label{sHx}
\end{eqnarray}
and $\check{H}_{xy}=\check{H}_{x\bar{y}}=0$.

For the $d_{x^2-y^2}$-wave SS,
\begin{eqnarray}
\check{H}_{0}=\left(\begin{array}{cc}
(\frac{4t_0}{a^2}-\mu)\sigma_{0}&0\\
0&(-\frac{4t_0}{a^2}+\mu)\sigma_{0}\end{array}\right),\label{d1H0}
\end{eqnarray}
\begin{eqnarray}
\check{H}_{x(y)}=\left(\begin{array}{cc}
-\frac{t_0}{a^2}\sigma_{0}&-(+)\frac{\Delta_{0}}{a^2}e^{i\phi_{l}} i\sigma_{y}\\
+(-)\frac{\Delta_{0}}{a^2}e^{-i\phi_{l}}i\sigma_{y} &\frac{t_0}{a^2}\sigma_{0}
\end{array}\right),\label{d1Hx}
\end{eqnarray}
and $\check{H}_{xy}=\check{H}_{x\bar{y}}=0$.

For the $d_{xy}$-wave SS, the matrix $\check{H}_0$ is the same with that in Eq.(\ref{d1H0}). The matrix $\check{H}_{x(y)}$ is the same with that in Eq.(\ref{sHx}). The matrix $\check{H}_{xy(x\bar{y})}$ is expressed as
\begin{eqnarray}
\check{H}_{xy(x\bar{y})}=\left(\begin{array}{cc}
0&-(+)\frac{\Delta_{0}}{2a^2}e^{i\phi_{l}}i\sigma_{y}\\
+(-)\frac{\Delta_{0}}{2a^2}e^{-i\phi_{l}}i\sigma_{y}&0
\end{array}\right).
\end{eqnarray}

For TS, the matrix $\check{H}_{0}$ is the same with that in Eq.(\ref{d1H0}). The matrices $\check{H}_{x}$ and $\check{H}_{y}$ are given by
\begin{eqnarray}
\check{H}_{x}=\left(\begin{array}{cc}
-\frac{t_0}{a^2}\sigma_{0}&\frac{-i\eta_1}{2a}\Delta_{0}e^{i\phi_{r}}\sigma_{x}\\
-\frac{i\eta_1}{2a}\Delta_{0}e^{-i\phi_{r}}\sigma_x&\frac{t_0}{a^2}\sigma_{0}
\end{array}\right),
\end{eqnarray}
and
\begin{eqnarray}
\check{H}_{y}=\left(\begin{array}{cc}
-\frac{t_0}{a^2}\sigma_{0}&\frac{\eta_2}{2a}\Delta_{0}e^{i\phi_{r}}\sigma_{x}\\
-\frac{\eta_2}{2a}\Delta_{0}e^{-i\phi_{r}}\sigma_x&\frac{t_0}{a^2}\sigma_{0}
\end{array}\right),
\end{eqnarray}
respectively.

For AM, the matrix $\check{H}_{0}$ is the same with that in Eq.(\ref{d1H0}). The matrix $\check{H}_{x(y)}=\text{diag}(h_{x(y)1},h_{x(y)2})$ with
$h_{x(y)1}=-\frac{t_0}{a^2}\sigma_0-(+)\frac{t_J\cos{2\alpha}}{a^2}\sigma_z$ and $h_{x(y)2}=\frac{t_0}{a^2}\sigma_0+(-)\frac{t_J\cos{2\alpha}}{a^2}\sigma_z$. The matrix $\check{H}_{xy(x\bar{y})}$ can be expressed as
\begin{eqnarray}
\check{H}_{xy(x\bar{y})}=\left(\begin{array}{cc}
-(+)\frac{t_J\sin{2\alpha}}{2a^2}\sigma_z&0\\
0&+(-)\frac{t_J\sin{2\alpha}}{2a^2}\sigma_z
\end{array}\right).
\end{eqnarray}

From the above matrices, the retard Green's function $G^{r}_{AM}(E,i_y,j_y)$ in AM and the free retard Green's function $g^{r}_{SS}(E,j_y,i_y)$ in SS can be obtained. Then, the lesser Green's function in Eq.(\ref{lG}) and the Josephson current in Eq.(\ref{Jc}) can be calculated.

\end{document}